# MEIGA
## Módulo Espacial con Integración de Giroscopio y Acelerómetro

Sistema de accesibilidad

# MEIGA
## Space module with gyroscope and accelerometer integration

Accessibility system

Dr. Antonio Losada González
anlosada@uvigo.es
Universidad de Vigo

1. Abstract

MEIGA is a module specially designed for people with tetraplegia or anyone who has very limited movement capacity in their upper limbs. MEIGA converts the user's head movements into mouse movements. To simulate keystrokes, it uses blinking, reading the movement of the cheek that occurs with it. The performance, speed of movement of the mouse and its precision are practically equivalent to their respective measurements using the hand.

2. Resumen

MEIGA es un módulo especialmente diseñado para personas con tetraplejia o cualquier persona que tenga muy limitada la capacidad de movimiento de sus miembros superiores. MEIGA convierte los movimientos de la cabeza del usuario en movimientos del ratón. Para simular la pulsación de teclas emplea el pestañeo, leyendo el desplazamiento de la mejilla que se produce con este. El rendimiento, la velocidad de desplazamiento del ratón y su precisión, son prácticamente equivalentes a sus respectivas medidas empleando la mano.

3. Introducción

MEIGA es un módulo sustitutorio del ratón que permite realizar todas sus funciones empleando movimientos de la cabeza. El sistema está especialmente diseñado para personas que han perdido la capacidad de mover con precisión sus miembros superiores pero conservan la capacidad de mover con precisión la cabeza empleando los movimientos del cuello. A diferencia de dispositivos anteriores, MEIGA es un dispositivo hardware que se monta anclado a la patilla de unas gafas. En caso de que el usuario necesite gafas, se anclará a las suyas propias, y en el caso de que no necesite, se ancla a una montura de gafas sin cristales. El dispositivo tiene un peso de 20 gr, por lo que apenas se hace notar y no consume potencia de procesamiento del propio ordenador. Una gran cantidad de dispositivos de seguimiento de cabeza para sustituir el ratón emplean cámaras web y un software de identificación y seguimiento facial [2][24][25]. Estos sistemas tienen como características positivas el emplear dispositivos estándar presentes en cualquier portátil, pero por otra banda consumen bastante potencia del procesador y presentan problemas cuando se desea mover el ratón con mucha precisión o a muy alta velocidad. Adicionalmente, si la potencia de procesamiento no es muy alta, presentan problemas para detectar las pulsaciones de ratón de alta velocidad.

Otros dispositivos emplean el seguimiento ocular [19][27]. Este tipo de dispositivos procesan las imágenes de los propios ojos o en el caso de estar iluminados por luz infrarroja, detectan el reflejo de la luz infrarroja en el fondo del ojo. Son sistemas bastante precisos, pero requieren el procesamiento de imágenes, además de dispositivos específicos de iluminación infrarroja. En este caso la pulsación se suele realizar mediante el pestañeo. Estos sistemas también presentan problemas con los movimientos de alta precisión, dado que necesitan captar movimientos minúsculos de los ojos. Otro inconveniente se presenta con el movimiento involuntario de los ojos al mirar alguna parte de la pantalla y dado su tecnología basada en el movimiento de los ojos, no permite independizar el movimiento del ratón de la zona que están mirando los ojos. Adicionalmente, hay que tener en cuenta que este sistema no puede ser aplicado en personas con nistagmus, dado que presentan movimientos involuntarios de los ojos.

También están alcanzando popularidad los sistemas de control del ratón mediante la detección de los movimientos de la lengua [6][14][16][26][29][30][31]. Estos sistemas son invasivos, dado que es necesario anclar un imán en la lengua del usuario, normalmente con un piercing. Alrededor de la cabeza se sitúan los detectores magnéticos. La lengua tiene capacidad de movimientos rápidos es un espacio 3D, lo que permite una alta sensibilidad.

En otros trabajos se ha experimentado con la detección de las señales eléctricas musculares (EMG-Electromiografía)[11][12]. Este sistema es no invasivo, requiriendo colocar bandas detectoras de impulsos eléctricos sobre los músculos que se pretenden leer. El problema de estos sistemas aplicados a personas con tetraplejia es que solo pueden ser empleados en los casos en que conservan alguna capacidad de movimiento muscular, aunque este no sea preciso.

Otra compleja línea de investigación son los sistemas de control BCI (Brain Computer Interface). Estos sistemas emplean cascos que se colocan sobre el cráneo y permiten leer las variaciones eléctricas de ciertas zonas del cerebro [8], llegando incluso a necesitar intervención quirúrgica para la implantación de los sensores en contacto con el cerebro [4][7].

Otros sistemas no invasivos permiten emplear la voz como sistema de control mediante aplicaciones que permiten procesar el sonido y transcribirlo a texto.[3][5][13][17][28].

Otros sistema de como [1] emplean la detección del soplido y la succión.

Como trabajos que utilizan sistemas muy poco empleados tenemos [17] que detecta el chasquido de los dientes mediante un sistema de detección de sonido anchado en la oreja.

Todos estos sistemas pueden ser empleados en módulos de control independiente, pero al combinarse en un solo sistema amplían la capacidad de control y gestión simultánea de aplicaciones, ejemplos de estos sistemas multifuncionales son [3][11][28].

4. Descripción

MEIGA es un pequeño dispositivo que porta una unidad inercial, que es la responsable de la captación de los movimientos de la cabeza de su usuario. Esta unidad posee un acelerómetro y un giróscopo con los que es capaz de detectar de modo muy preciso tanto los movimientos de cabeceo, alabeo como guiñada. Estos términos son muy empleados en la jerga de aviación y su significado puede apreciarse de modo muy gráfico en la Imagen 1.

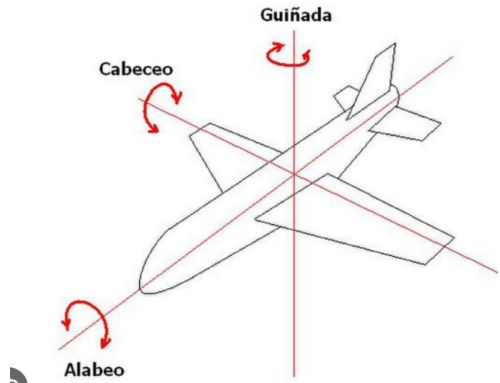
*Imagen 1. Movimientos de control de MEIGA.*

Integrando los valores de velocidad angular es capaz de detectar las ligeras variaciones en la posición de la cabeza del usuario.

La sustitución de los botones se realiza mediante el pestañeo y el movimiento de la mejilla que se produce con él. Mediante un pestañeo rápido se genera la pulsación del botón izquierdo, con un pestañeo superior a 1 segundo se genera la pulsación del botón derecho y con pestañeo de duración mayor se pulsa el botón izquierdo de modo continuo para seleccionar texto, mover ventanas, etc.

El dispositivo se alimenta por USB o batería conectada al puerto USB-C de MEIGA. Dependiendo del dispositivo con el que se quiere usar MEIGA, se conectará a un teléfono, ordenador o Tablet. La conexión con el ordenador, teléfono o tableta la realiza por bluetooth. MEIGA se presenta como un dispositivo de entrada HID de tipo ratón, por lo que se reconoce automáticamente y se instala sin necesidad de aplicaciones. Es compatible tanto con Android, Windows y iPhone. Una vez conectado hay que emparejarlo por bluetooth.

El dispositivo se ancha a cualquier tipo de gafa, con cuidado de que el sensor de pestañeo se encuentre en la zona de movimiento de la mejilla. El dispositivo se autoconfigura en el arranque. Una vez conectado, el sistema lee la posición de la mejilla en reposo para calcular la variación de la posición base que generará las pulsaciones del ratón.

En la Imagen 2 se puede ver un módulo MEIGA montado sobre la pata de una montura de unas gafas.

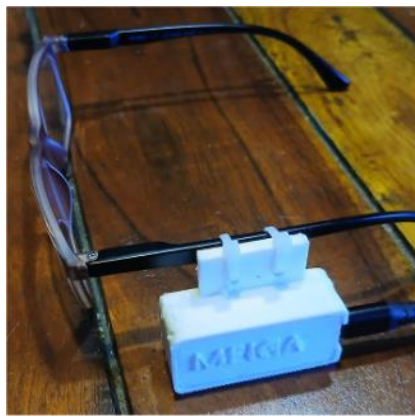
*Imagen 2. Montaje de MEIGA sobre la montura de las gafas.*

El sistema permite simular las siguientes funciones:
1. Movimiento vertical del ratón
2. Movimiento horizontal del ratón
3. Desplazamiento de pantalla hacia arriba y hacia abajo
4. Pulsación corta del botón derecho
5. Pulsación corta del botón izquierdo
6. Pulsación corta del botón central
7. Pulsación mantenida del botón izquierda

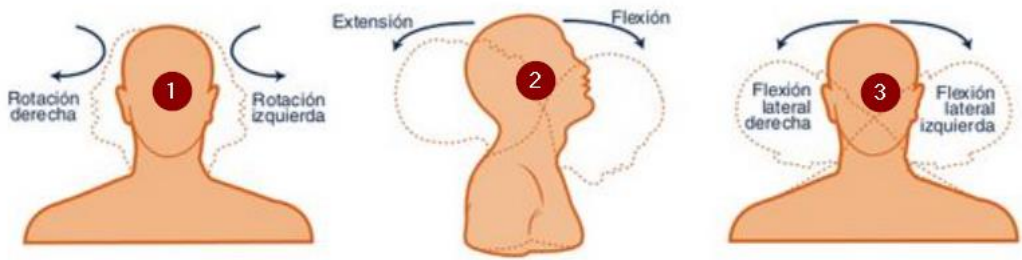

*Imagen 3. Movimientos de control del ratón.*

En la Imagen 3 puede apreciarse gráficamente los movimientos de cabeza que permiten realizar el control del ratón. Con (1) está identificado el movimiento de guiñada que mueve el ratón lateralmente de izquierda a derecha y viceversa en el eje horizontal. Con (2) se identifica el movimiento de cabeceo que desplaza el ratón hacia arriba y hacia abajo en el eje vertical y con (3) se identifica el movimiento de alabeo que permite realizar un desplazamiento del contenido de la pantalla hacia arriba y hacia abajo.

5. Esquema de componentes

MEIGA está compuesta por 3 componentes de hardware. Estos componentes han sido seleccionados para mantener el coste del dispositivo lo más bajo posible permitiendo que se pueda montar en un espacio reducido con un peso muy contenido, dado que debe anclarse a una montura de gafas.

Los componentes son los siguientes:

1. ESP32-C3 Super Mini. Es el procesador principal
2. MPU6050. Es una unidad de medición inercial. Recupera la posición de la cabeza del usuario.
3. TCRT5000. Se encarga de leer la posición de las mejillas para detectar el pestañeo.

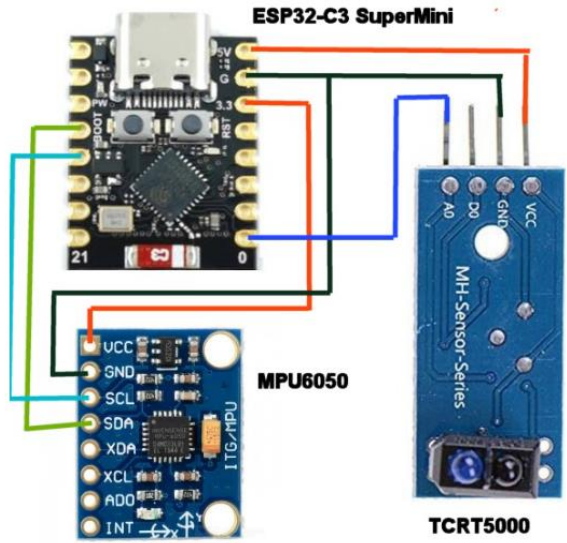
*Imagen 4. Esquema de conexiones de MEIGA.*

En la Imagen 4 se muestran los componentes y las conexiones entre ellos. Dado que se ha querido mantener su tamaño lo más reducido posible, no se ha diseñado una placa de conexiones adicionales, y estas se realizan soldando los cables de unión. Adicionalmente se han escogido módulos muy económicas y fácilmente disponibles en minoritas de venta de electrónica, dado que el objetivo es publicar el sistema como un módulo "open source" para que cualquier organización puedan montarlo y ponerlo a disposición de sus usuarios. De hecho, el dispositivo ha sido desarrollado a petición del Instituto Novo Ser de Rio de Janeiro. Este instituto nos solicitó un dispositivo económico y de alta precisión para el control del ratón que pudiera ser empleado por personas con tetraplejia. Adicionalmente nos pidió que el dispositivo pudiera ser construido por personas sin conocimientos técnicos con las instrucciones adecuadas.

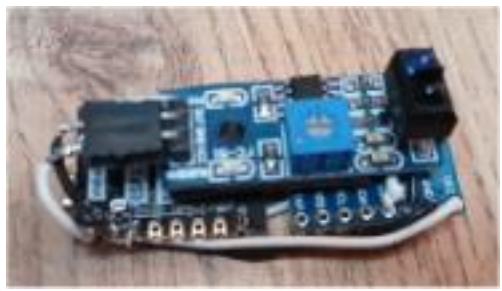
*Imagen 5. Posición de montaje de los componentes de hardware.*

El resultado final del hardware montado es el siguiente, en donde se puede apreciar el módulo de medición de distancia de la mejilla en la parte superior (TCRT5000) y debajo a la derecha la unidad de medida inercial (MPU6050) y en la parte izquierda el microcontrolador (ESP32-C3 Super Mini).

6. Diseño 3D de la caja

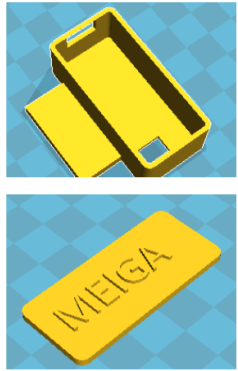

*Imagen 6. Diseño 3D de la caja de MEIGA.*

Para contener el dispositivo ha sido necesario diseñar una caja (Imagen 6) que se ha puesto a disposición de la comunidad tanto en formato fuente FreeCad como STL en la siguiente dirección web:

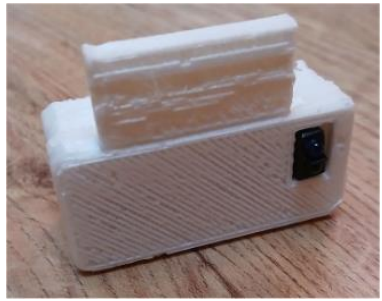

*Imagen 7. Imagen real de MEIGA.*

Una vez impresa, el resultado se puede ver en la Imagen 7, en donde podemos ver la lengüeta que sobresale en la parte superior para anclarla a la patilla de las gafas, junto con la ventana en donde se encuentra el emisor y receptor infrarrojo del módulo TCRT5000 encargado de la lectura de distancia a la mejilla.

7. Lectura de información de la unidad inercial

La unidad inercial está compuesta por un acelerómetro y un giróscopo. El acelerómetro devuelve la aceleración instantánea y el giróscopo devuelve los radianes por segundos de giro.

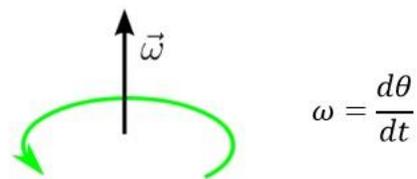

$$\omega = \frac{d\theta}{dt}$$

*Imagen 8. Rotación del giróscopo.*

La unidad se sitúa en la caja en posición vertical.

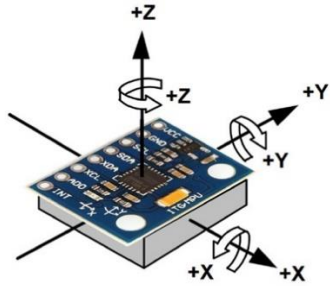

*Imagen 9. Unidad IMU MPU6050.*

La MPU6050 posee un sensor MEMS, cuyo ligero desplazamiento genera corrientes eléctricas que pueden ser medidas.

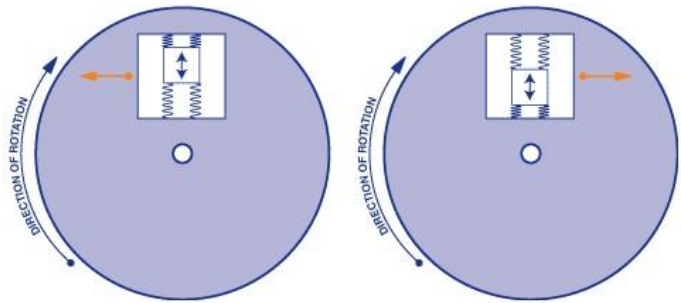

*Imagen 10. Funcionamiento de un sensor MEMS.[1]*

El giroscopio contiene un pequeño sensor MEMS (de entre 1 y 100 micrómetros). La masa del sensor MEMS en movimiento genera pequeñas corrientes eléctricas que son amplificadas para que puedan ser leídas por el microcontrolador. La unidad inercial recupera la información del acelerómetro y del giróscopo. La medida del acelerómetro es absoluta, obteniendo la aceleración en los tres ejes. Estos valores contienen un pequeño error. La medida del giróscopo se recupera como una unidad de velocidad rotación (radianes/segundo). Para calcular la posición es necesario integrar las sucesivas medidas para obtener la posición actual. La medida del giróscopo es muy precisa, pero debido a que es necesario acumular los cientos de medidas que se van recuperando en el tiempo, la posición final va acumulando errores que al final resultan en errores importantes.

Dado que la medida del acelerómetro es absoluta y la medida del giróscopo es incremental o relativa a la posición de la última lectura, los sistemas de cálculo de posición emplean ambas medidas para obtener la posición con la mayor precisión posible. Para combinar ambas medidas podemos emplear distintos filtros, entre los más conocidos podemos citar el filtro complementario por su sencillez y el filtro de Kalman por su precisión.

---

[1] https://uvadoc.uva.es/bitstream/handle/10324/12884/TFG-P161.pdf;jsessionid=2CDB11E7DE1C0F4B39DF708888257868?sequence=1

8. Filtro complementario

El filtro complementario combina ambas medidas ponderadas con las constantes $A$ y $B$. Normalmente se emplean valores como $A = 0.95$ y $B = 0.05$. Con estos valores, a corto plazo se da mucho valor a las medidas del giróscopo y este valor se va corrigiendo con las medidas del acelerómetro con el paso del tiempo, lo que evita la deriva.

$$\theta = A \cdot (\theta_{prev} + \theta_{giro}) + B \cdot \theta_{acel}$$

9. Filtro de Kalman

El filtro de Kalman es un algoritmo que permite estimar variables ocultas, no observables, a partir de variables que pueden ser medidas con cierto error.

La formulación para sistemas estáticos del filtro de Kalman se muestra en las Ecuaciones (1) a (5). Inicialmente SAPPO está diseñado para ser empleado en sistemas de navegación de robots, pero dado que puede ser empleado en robots con cualquier formulación cinemática, el filtro se empleará únicamente para eliminar el ruido de las medidas de distancia de las balizas.

$$\hat{x}_{n,n} = \hat{x}_{n,n-1} + K_n(z_n - \hat{x}_{n,n-1}) \tag{1}$$
$$\hat{x}_{n+1,n} = \hat{x}_{n,n} \tag{2}$$
$$K_n = \frac{P_{n,n-1}}{P_{n,n-1} - r_n} \tag{3}$$
$$P_{n,n} = (1 - K_n)P_{n,n-1} \tag{4}$$
$$P_{n+1,n} = P_{n,n} + q_n \tag{5}$$

El filtro tiene los siguientes parámetros: en la Ecuación (1), $z_n$ es el resultado de la medición, en la Ecuación (3), $r_n$ es la varianza del error de medición y en la Ecuación (5), $q_n$ es la varianza del error del proceso, ambas varianzas deben seguir una distribución normal del media cero.

La Ecuación (1) obtiene el estado actual del sistema y la Ecuación (2) realiza la predicción obteniendo el estado estimado futuro. Dado que emplearemos el filtro para eliminar el ruido en las medidas, no aplicaremos ninguna formulación para predecir el estado futuro, por lo que el estado futuro se iguala al estado actual.

En cada iteración, el algoritmo de Kalman tiene que calcular el estado actual del proceso para lo que emplea la Ecuación (1). Esta ecuación necesita el estado anterior $\hat{x}_{n,n-1}$, junto con la medida $z_n$ y el valor de la ganancia de Kalman $K_n$. Los dos primeros valores son conocidos por lo que debe calcularse la ganancia de Kalman con la Ecuación (3) que necesita la varianza del error del proceso calculada en la iteración anterior $P_{n,n-1}$, junto con la varianza del error de estimación de la medida $r_n$, ambos conocidos. La predicción del estado siguiente se realiza mediante la Ecuación (2) que, dado que se está empleando el filtro para eliminar el ruido de las medidas, no tenemos la formulación de la estimación dinámica del sistema, por lo que la predicción del estado siguiente se iguala al estado actual. Finalmente se actualiza el valor de la covarianza del error del proceso $P_{n,n}$ con la Ecuación (4) para finalmente calcular la predicción de la covarianza del error del proceso $P_{n+1,n}$ en la Ecuación (5).

Una vez calculada la predicción, debe ajustarse la ganancia de Kalman, tarea que será realizada por la Ecuación (3). Para calcular la ganancia de Kalman se empleará la información de la covarianza en el instante de tiempo anterior $P_{n,n-1}$. Una vez calculada la ganancia $K_n$, se procede a actualizar el valor de la covarianza en la Ecuación (4) para finalmente estimar la covarianza futura en la Ecuación (5).

10. Sistema de detección del pestañeo

Para la detección del pestañeo se emplea un sensor TCRT5000. Este sensor es capaz de medir con precisión la distancia a un objeto. En este caso debe orientarse el sensor hacia la mejilla para que sea capaz de detectar el movimiento de la mejilla que se genera al pestañear. Con cierto entrenamiento es posible mover la mejilla sin necesidad de cerrar los párpados, con lo que podremos hacer click del ratón con mayor precisión

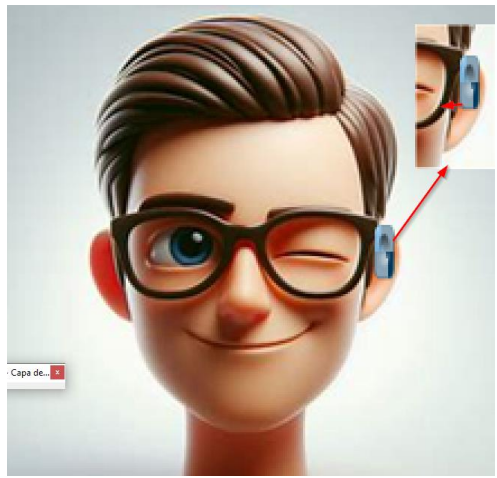

*Imagen 11.Click del ratón mediante el pestañeo.*

En la Imagen 11 se puede ver el dispositivo MEIGA y el lugar donde se encuentra en sensor de distancia apuntando a la mejilla.

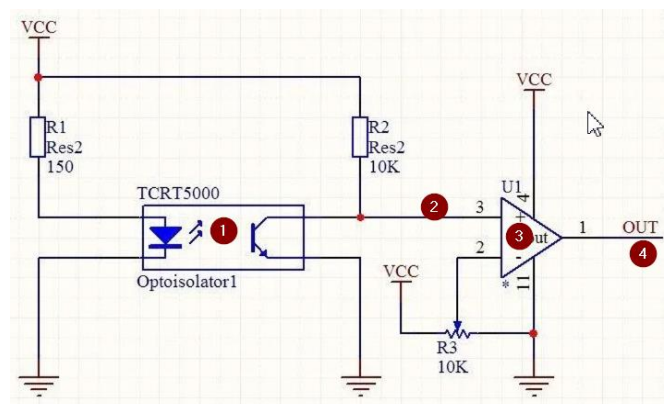

*Imagen 12.Esquema electrónico do módulo TCRT5000.*

Los módulos TCRT5000 que se pueden adquirir en tiendas minoristas son módulos integrales con salida digital. Normalmente estos módulos son empleados como módulos digitales que son capaces de activarse en el caso de que un objeto se encuentre dentro del rango configurado. El rango o distancia de detección del objeto se parametriza con un potenciómetro cuya salida se toma como entrada del comparador marcado como (3) en la Imagen 12 para obtener la salida digital (4). Adicionalmente es habitual que dispongan una salida analógica conectada a la patilla marcada como (2) en la Imagen 12. Esta patilla devuelve un potencial entre 0 y 5v dependiendo de la distancia a la que se encuentre el objeto.

## 11. Características del sensor de detección del pestañeo

El sensor TCRT5000 tiene un coste extremadamente bajo, aunque presente varios problemas. Es un sensor de lectura por reflexión. Está compuesto por un emisor de luz infrarroja y un detector de luz reflejada por el objeto (en este caso la mejilla). La cantidad de luz reflejada será proporcional a la distancia de la mejilla, pero adicionalmente, esta medida puede verse modificada por otros factores como la textura o el color de la piel, con colores oscuros, la cantidad de luz reflejada puede ser muy bajo. Incluso puede estar afectado por otras fuentes de emisión de luz infrarroja como puede ser un monitor, siendo posible que esta afectación sea variable y dependa de la posición de la cabeza. Sería deseable disponer de un sensor de medición de distancia por láser, pero el problema de estos sensores se encuentra en el rango de medición mínimo que suele ser del rango de varios centímetros, cuando la distancia del sensor a la mejilla está en el rango de milímetros.

## 12. Referencias

[1] H.-C. Chen, C.-J. Huang, W.-R. Tsai, y C.-L. Hsieh, «A computer mouse using blowing sensors intended for people with disabilities», *Sensors*, vol. 19, n.º 21, p. 4638, 2019.

[2] Y.-L. Chen, W.-L. Chen, T.-S. Kuo, y J.-S. Lai, «A head movement image (HMI)-controlled computer mouse for people with disabilitiesAnalysis of a time-out protocol and its applications in a single server environment», *Disability and Rehabilitation*, vol. 25, n.º 3, pp. 163-167, ene. 2003, doi: [10.1080/0963828021000024960](10.1080/0963828021000024960).

[3] M. N. Sahadat, A. Alreja, P. Srikrishnan, y M. Ghovanloo, «A multimodal human computer interface combining head movement, speech and tongue motion for people with severe disabilities», en *2015 IEEE biomedical circuits and systems conference (BioCAS)*, IEEE, 2015, pp. 1-4. Accedido: 31 de agosto de 2024. [En línea]. Disponible en: https://ieeexplore.ieee.org/abstract/document/7348317/

[4] B. Jarosiewicz *et al.*, «Advantages of closed-loop calibration in intracortical brain–computer interfaces for people with tetraplegia», *Journal of neural engineering*, vol. 10, n.º 4, p. 046012, 2013.

[5] M. Soares, L. Mesquita, F. Oliveira, y L. Rodrigues, «ANA: A Natural Language System with Multimodal Interaction for People Who Have Tetraplegia», en *Universal Access in Human-Computer Interaction. Multimodality and Assistive Environments*, vol. 11573, M. Antona y C. Stephanidis, Eds., en Lecture Notes in Computer Science, vol. 11573. , Cham: Springer International Publishing, 2019, pp. 353-362. doi: [10.1007/978-3-030-23563-5_28](10.1007/978-3-030-23563-5_28).

[6] J. Kim *et al.*, «Assessment of the tongue-drive system using a computer, a smartphone, and a powered-wheelchair by people with tetraplegia», *IEEE Transactions on Neural Systems and Rehabilitation Engineering*, vol. 24, n.º 1, pp. 68-78, 2015.

[7] J. P. Donoghue, A. Nurmikko, M. Black, y L. R. Hochberg, «Assistive technology and robotic control using motor cortex ensemble-based neural interface systems in humans with tetraplegia», *The Journal of Physiology*, vol. 579, n.º 3, pp. 603-611, mar. 2007, doi: 10.1113/jphysiol.2006.127209.

[8] A.-L. Benabid *et al.*, «Brain Computer Interface Driven Neuroprosthesis for Tetraplegic Patients: Preliminary Steps in Rats and Primates: 966», *Neurosurgery*, vol. 67, n.º 2, pp. 557-558, 2010.

[9] T. Felzer, I. S. MacKenzie, y J. Magee, «Comparison of Two Methods to Control the Mouse Using a Keypad», en *Computers Helping People with Special Needs*, vol. 9759, K. Miesenberger, C. Bühler, y P. Penaz, Eds., en Lecture Notes in Computer Science, vol. 9759. , Cham: Springer International Publishing, 2016, pp. 511-518. doi: 10.1007/978-3-319-41267-2_72.

[10] B. Delhaye, M. Schramme, P. Doguet, J.-D. Legat, y F.-X. STANDAERT, «Design of a BMI for tetraplegic patients», Accedido: 31 de agosto de 2024. [En línea]. Disponible en: https://dial.uclouvain.be/downloader/downloader.php?pid=thesis%3A6738&datastream=PDF_01

[11] G. Lee, K. Kim, y J. Kim, «Development of hands-free wheelchair device based on head movement and bio-signal for quadriplegic patients», *International Journal of Precision Engineering and Manufacturing*, vol. 17, pp. 363-369, 2016.

[12] L. Kauhanen, P. Jylänki, J. Lehtonen, P. Rantanen, H. Alaranta, y M. Sams, «EEG-Based Brain-Computer Interface for Tetraplegics», *Computational Intelligence and Neuroscience*, vol. 2007, pp. 1-11, 2007, doi: 10.1155/2007/23864.

[13] S. Pouplin *et al.*, «Effect of a dynamic keyboard and word prediction systems on text input speed in patients with functional tetraplegia», *Journal of rehabilitation research and development*, vol. 51, n.º 3, pp. 467-480, 2014.

[14] N. Sebkhi *et al.*, «Evaluation of a head-tongue controller for power wheelchair driving by people with quadriplegia», *IEEE Transactions on Biomedical Engineering*, vol. 69, n.º 4, pp. 1302-1309, 2021.

[15] N. Rudigkeit, M. Gebhard, y A. Gräser, «Evaluation of control modes for head motion-based control with motion sensors», en *2015 IEEE International Symposium on Medical Measurements and Applications (MeMeA) Proceedings*, IEEE, 2015, pp. 135-140. Accedido: 31 de agosto de 2024. [En línea]. Disponible en: https://ieeexplore.ieee.org/abstract/document/7145187/

[16] N. Sebkhi *et al.*, «Evaluation of Multimodal Tongue Drive System by People with Tetraplegia for Computer Access», 9 de abril de 2021. doi: 10.36227/techrxiv.14388113.v1.

[17] T. Simpson, M. Gauthier, y A. Prochazka, «Evaluation of Tooth-Click Triggering and Speech Recognition in Assistive Technology for Computer Access», *Neurorehabil Neural Repair*, vol. 24, n.º 2, pp. 188-194, feb. 2010, doi: 10.1177/1545968309341647.

[18] A. Folan, L. Barclay, C. Cooper, y M. Robinson, «Exploring the experience of clients with tetraplegia utilizing assistive technology for computer access», *Disability and Rehabilitation: Assistive Technology*, vol. 10, n.º 1, pp. 46-52, ene. 2015, doi: 10.3109/17483107.2013.836686.

[19] M. Caligari, M. Godi, S. Guglielmetti, F. Franchignoni, y A. Nardone, «Eye tracking communication devices in amyotrophic lateral sclerosis: Impact on disability and quality of life», *Amyotrophic Lateral Sclerosis and Frontotemporal Degeneration*, vol. 14, n.º 7-8, pp. 546-552, dic. 2013, doi: 10.3109/21678421.2013.803576.

[20] K. Sancheti, A. Suhaas, y P. Suresh, «Hands-free cursor control using intuitive head movements and cheek muscle twitches», en *TENCON 2018-2018 IEEE Region 10 Conference*, IEEE, 2018, pp. 0356-0361. Accedido: 31 de agosto de 2024. [En línea]. Disponible en: https://ieeexplore.ieee.org/abstract/document/8650532/?casa_token=qVMFanELAqUAAAAA:JX3Sw3oRtr_olJF21GdmgD05DxOeF2W45xIdmSgosaaEDSLou7wpx0KGkv-IW0k5e7Sau8w

[21] A. Jackowski, M. Gebhard, y R. Thietje, «Head Motion and Head Gesture Based Robot Control for Tetraplegics: A Usability Study», Accedido: 31 de agosto de 2024. [En línea]. Disponible en: https://www.en.w-hs.de/fileadmin/Oeffentlich/Fachbereich-2/PT/Lehrbereiche/SuA/Gebhard_Veroeffentlichungen/Journal_Paper_AJ_2017-02-24_final.pdf

[22] A. HeydariGorji, S. M. Safavi, C. T. Lee, y P. H. Chou, «Head-mouse: A simple cursor controller based on optical measurement of head tilt», en *2017 IEEE Signal Processing in Medicine and Biology Symposium (SPMB)*, IEEE, 2017, pp. 1-5. Accedido: 31 de agosto de 2024. [En línea]. Disponible en: https://ieeexplore.ieee.org/abstract/document/8257058/?casa_token=nG62ZrAcY40AAAAA:6bl1q4jGWxTUjwQxrb38HlRMXrdL1ix1O7GoDNX6spy61H_RhEQGJAyLfnPPZIazL51Cos4

[23] J. D. Simeral *et al.*, «Home use of a wireless intracortical brain-computer interface by individuals with tetraplegia», *medRxiv*, pp. 2019-12, 2019.

[24] R. Shashidhar, K. Snehith, P. K. Abhishek, A. B. Vishwagna, y N. Pavitha, «Mouse Cusor Control Using Facial Movements-An HCI Application», en *2022 International Conference on Sustainable Computing and Data Communication Systems (ICSCDS)*, IEEE, 2022, pp. 367-371. Accedido: 31 de agosto de 2024. [En línea]. Disponible en: https://ieeexplore.ieee.org/abstract/document/9760914/?casa_token=GQFTMJboYXIAAAAA:jiB93gLvrXfzx_qJBiuLzIUJas3a2v3aAYIuFJJJ7tMLWfmyqg5_k6DguLotulqq43JiJ78

[25] S. S. Khan, M. S. H. Sunny, M. S. Hossain, E. Hossain, y M. Ahmad, «Nose tracking cursor control for the people with disabilities: An improved HCI», en *2017 3rd International Conference on Electrical Information and Communication Technology (EICT)*, IEEE, 2017, pp. 1-5. Accedido: 31 de agosto de 2024. [En línea]. Disponible en: https://ieeexplore.ieee.org/abstract/document/8275178/?casa_token=B-1X7UFj_2gAAAAA:gzzAWqGL1K-hKfmkDD98taeOlKwLS_CtHDGteLspnrYtKhfVmBi1NbOCUH_IBK4hfURw4LU

[26] M. N. Sahadat, N. Sebkhi, D. Anderson, y M. Ghovanloo, «Optimization of tongue gesture processing algorithm for standalone multimodal tongue drive system», *IEEE Sensors Journal*, vol. 19, n.º 7, pp. 2704-2712, 2018.

[27] S. Stalljann, L. Wöhle, J. Schäfer, y M. Gebhard, «Performance analysis of a head and eye motion-based control interface for assistive robots», *Sensors*, vol. 20, n.º 24, p. 7162, 2020.

[28] M. N. Sahadat, A. Alreja, y M. Ghovanloo, «Simultaneous multimodal PC access for people with disabilities by integrating head tracking, speech recognition, and tongue motion», *IEEE transactions on biomedical circuits and systems*, vol. 12, n.º 1, pp. 192-201, 2017.

[29] J. Kim *et al.*, «The Tongue Enables Computer and Wheelchair Control for People with Spinal Cord Injury», *Sci. Transl. Med.*, vol. 5, n.º 213, nov. 2013, doi: 10.1126/scitranslmed.3006296.

[30] E. R. Lontis, B. Bentsen, M. Gaihede, F. Biering-Sørensen, y L. N. A. Struijk, «Wheelchair control with inductive intra-oral tongue interface for individuals with tetraplegia», *IEEE Sensors Journal*, vol. 21, n.º 20, pp. 22878-22890, 2021.

[31] L. N. S. Andreasen Struijk, L. L. Egsgaard, R. Lontis, M. Gaihede, y B. Bentsen, «Wireless intraoral tongue control of an assistive robotic arm for individuals with tetraplegia», *J NeuroEngineering Rehabil*, vol. 14, n.º 1, p. 110, dic. 2017, doi: 10.1186/s12984-017-0330-2.